\documentclass[a4paper,10pt,twoside]{cpc-hepnp}

\usepackage{url}
\usepackage{multicol}
\usepackage{graphicx}
\usepackage{booktabs}
\usepackage{amssymb,bm,mathrsfs,bbm,amscd}
\usepackage[tbtags]{amsmath}
\usepackage{lastpage}
\usepackage{CJK}
\usepackage{lineno}

\usepackage[abs]{overpic}

\makeatletter
    
    \newcommand{\Rmnum}[1]{\expandafter\@slowromancap\romannumeral #1@}
\makeatother

\begin{document}

\begin{CJK*}{GBK}{song}

\fancyhead[c]{\small Submitted to Chinese Physics C~~}
\fancyfoot[C]{\small \thepage}

%\footnotetext[0]{Received X Y 2015}
%, Revised dd mm 2015}

\title{Production of $\gamma\gamma$ $+$ 2 jets from double parton scattering in proton-proton collisions at the LHC
\thanks{Supported by National Natural
Science Foundation of China (No. 11061140514), China Ministry of
Science and Technology (No. 2013CB838700) and in part by the CAS
Center for Excellence in Particle Physics (CCEPP)}}

\author{
\quad TAO Jun-Quan$^{1;1)}$\email{taojq@mail.ihep.ac.cn}%%
\quad ZHANG Si-Jing$^{1,2}$
\quad SHEN Yu-Qiao$^{1,2}$ \\
\quad FAN Jia-Wei$^{1,2}$
\quad CHEN Guo-Ming$^{1}$
\quad CHEN He-Sheng$^{1}$  \\
%\quad X $^{1}$  \quad Y $^{1}$  \quad Z $^{1}$
}

\maketitle

\address{
$^1$ Institute of High Energy Physics, Chinese Academy of Sciences, Beijing 100049, China \\
\vspace*{4pt}
$^2$ University of Chinese Academy of Sciences, Beijing 100049, China \\

}

\begin{abstract}
Cross sections for the production of pairs of photons plus two
additional jets produced from double parton scattering in
high-energy proton-proton collisions at the LHC are calculated for
the first time. The estimates are based on the theoretical
perturbative QCD predictions for the productions of $\gamma\gamma$
at next-to-next-to-leading-order, jet $+$ jet and $\gamma$ $+$ jet
at next-to-leading-order, for their corresponding single-scattering
cross sections.
%These theoretical predictions for these three
%processes can give the best agreements with the measurements from
%the LHC data in years 2011 and 2012, compared to other theoretical
%predictions.
The cross sections and expected event rates for
$\gamma\gamma$ $+$ 2 jets from double parton scattering, after
typical acceptance and selections, are given for proton-proton
collisions with the collision energy $\sqrt{s}$ = 13~TeV and an
integrated luminosity of 100 $fb^{-1}$ planned for the following two
years, and also $\sqrt{s}$ = 14~TeV with 3000 $fb^{-1}$ of
integrated luminosity as LHC designed.

\end{abstract}



%\listset{numbers=left, numberstyle=\bf, stepnumber=1, numbersep=20pt}

%\begin{linenumbers}

\begin{multicols}{2}

\section{Introduction}

In proton-proton (pp) collisions with higher energies at the Large
Hadron Collider (LHC), particle production is dominated by multiple
interactions of their constituent partons, with most particles from
the hardest proton-proton scattering and the radiation and
fragmentation of secondary partonic actions. The higher
centre-of-mass energy leads to enhanced parton densities which cause
a sizable probability for two or more parton-parton scatterings
within the same pp interaction
\cite{labDPSTheory1,Lansberg:2014swa}. At LHC, various measurements
of the differential distributions in W + jets
\cite{labATLASW2Jets,labCMSW2Jets} and $J/\psi$\ + W
\cite{labATLASWJPsi} show that the excesses above the expectations
from single parton scattering (SPS) are consistent with double
parton scattering (DPS). Various measurements in other pp and
$p\bar{p}$ collisions at $\sqrt{s}$ = 63~GeV \cite{labUA2MultiJets},
630~GeV \cite{labAFSMultiJets}, and 1.8~TeV \cite{labCDFMultiJets}
are consistent with DPS contributions to multijet final states, as
well as to $\gamma$ + 3-jet events at $\sqrt{s}$ = 1.8~TeV
\cite{labCDFGamma3Jets} and 1.96 ~TeV \cite{labD0Gamma3Jets}. The
measurements of DPS processes can provide valuable information on
the transverse distribution of partons in the proton
\cite{labDPSTheoryMultiQCD} and on the multi-parton correlations in
the hadronic wave function \cite{labDPSTheoryHWF}. DPS also
constitutes the background for new physics searches at the LHC
\cite{labDPSBkg1,labDPSBkg2,labDPSBkg3}. Additional searches for DPS
have been proposed via double Drell-Yan, four jets and same-sign WW
production \cite{labDPSDY,labDPS4Jets,labDPSWW}.

In this paper, the cross section for the production of pairs of
photons plus two additional jets produced from double parton
scattering (DPS) in high-energy proton-proton collisions at the LHC
are calculated for the first time. $\gamma\gamma$ final states have
played a crucial role in the recent discovery of a new boson at the
LHC \cite{labCMSHiggs,labATLASHiggs} and are also important in many
New Physics searches
\cite{labATLASDiphotonMET,labATLASDiphotonED,labCMSDiphotonMET,labCMSDiphotonNP},
in particular the search for extra spatial dimensions or cascade
decays of heavy new particles. In particular, diphotons in
combination with jets and missing energy occur in gauge mediated
SUSY scenarios. $\gamma\gamma$ or plus two additional jets offers
also an important test of both perturbative and non-perturbative
quantum chromodynamics (QCD)
\cite{labCMSDiphoton7TeV,labJTaoDiphoton7TeV,labATLASDiphoton7TeV}.
$\gamma\gamma$ $+$ 2 jets is also the main irreducible background
for other physics analyses with $\gamma\gamma$ and jets in the final
state at the LHC, such as Higgs produced in vector boson fusion. For
the production of $\gamma\gamma$ $+$ 2 jets, a sizeable contribution
from DPS with $\gamma\gamma$ produce in one scattering while the
second scattering yielding two jets can be expected.

The structure of this paper is organized as follows. In section 2, a
generic way of the DPS cross section as the product of the SPS cross
sections and its parameter are briefly introduced. The details of
the cross section of $\gamma\gamma$ $+$ 2 jets calculation and the
cross section of different SPS processes estimated from higher order
theoretical predictions are described in section 3. The results
including the cross section of $\gamma\gamma$ $+$ 2 jets with
$\sqrt{s}$ = 13~TeV and 14~TeV at LHC and the expected event rates,
with typical selections,a re summarized in section 4. The summary
and outlook are given in section 5.

%\section{Generic description of DPS}
\section{Generic formule of DPS}

For a composite system ($A+B$) in hadronic collisions, its
production cross section from DPS, $\sigma_{pp \to AB}^{DPS}$, can
be written model-independently as the product of the cross sections
of $A$ and $B$ originated from single parton scattering, $\sigma_{pp
\to A}^{SPS}$ and $\sigma_{pp \to B}^{SPS}$, normalized by an
effective cross section $\sigma_{eff}$ \cite{labDPSgeneral}

\begin{equation}
\begin{split}
\label{eqGeneralDPS} \sigma_{pp \to AB}^{DPS} =
\frac{m}{2}\frac{\sigma_{pp \to A}^{SPS}\times \sigma_{pp \to
B}^{SPS}}{\sigma_{eff}},
\end{split}
\end{equation}
where $m$ ia a symmetry factor accounting for distinguishable
($m$=2) and indistinguishable ($m$=1) final-states.

The effective cross section $\sigma_{eff}$ is a measure of the
transverse distribution of partons inside the colliding hadrons and
their overlap in a collision. It is independent of the process and
of the phase-space under consideration. A number of measurements of
$\sigma_{eff}$  have been performed in pp and $p\bar{p}$ collisions
at $\sqrt{s}$ = 63~GeV \cite{labUA2MultiJets}, 630~GeV
\cite{labAFSMultiJets}, 1.8~TeV
\cite{labCDFMultiJets,labCDFGamma3Jets,labCDFGamma3JetsCorr}, 1.96
~TeV \cite{labD0Gamma3Jets} and also 7~TeV at LHC
\cite{labATLASW2Jets,labCMSW2Jets}. The measured values range from
5mb at the lowest energy to about 20mb from CMS at 7~TeV.
Fig.~\ref{fig:effsigma} shows a comparison of the effective cross
section $\sigma_{eff}$ measured by different experiments using
different processes at various centre-of-mass energies.

\begin{center}
\includegraphics[width=7cm]{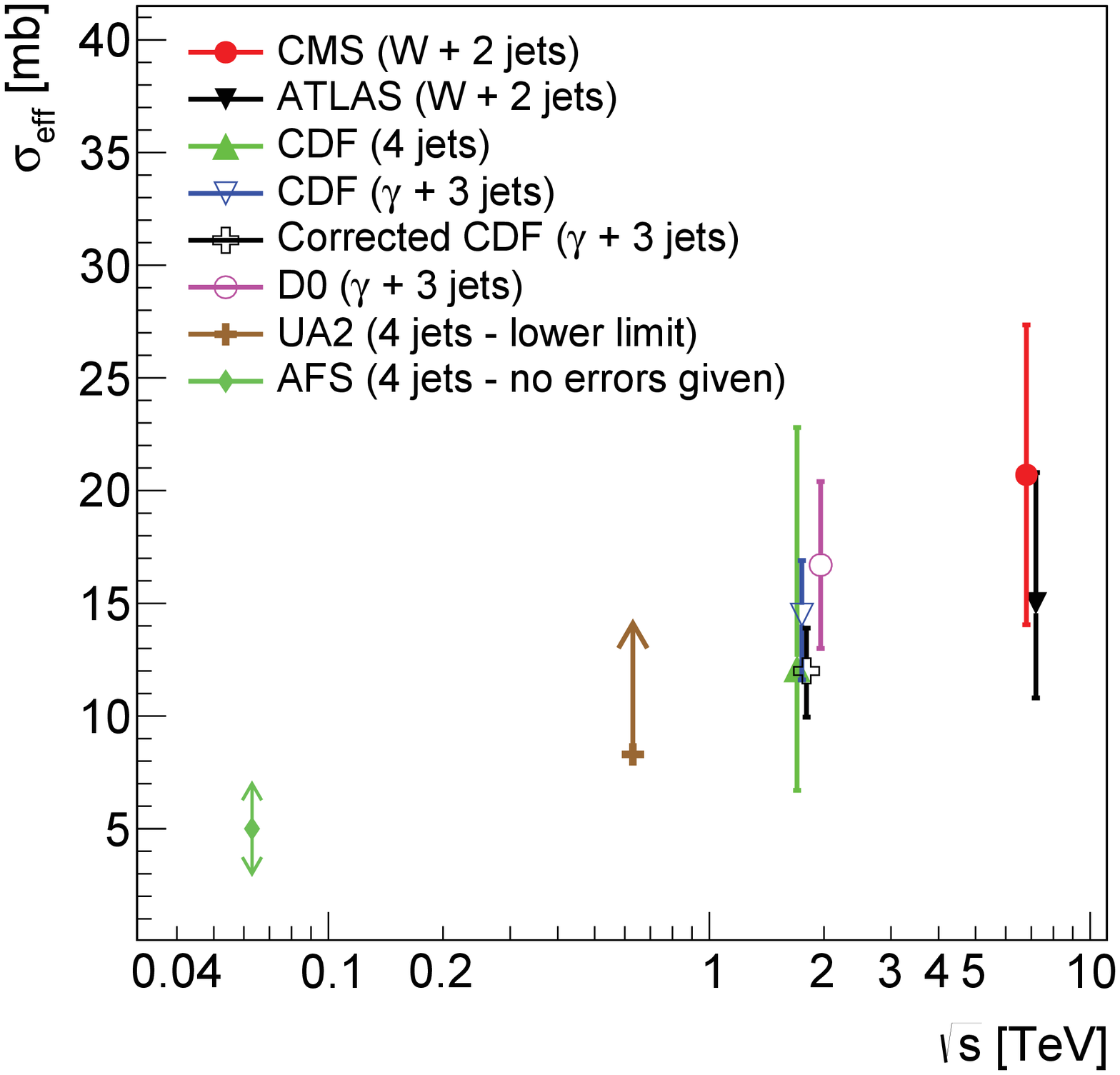}
 \figcaption{ $\sigma_{eff}$ measured by different experiments using different processes
\cite{labUA2MultiJets,labAFSMultiJets,labCDFMultiJets,labCDFGamma3Jets,labD0Gamma3Jets,labATLASW2Jets,labCMSW2Jets,labCDFGamma3JetsCorr}.
%Different methods were used for the extraction of the DPS fraction
%and $\sigma_{eff}$.
The "Corrected CDF" data point indicate the
$\sigma_{eff}$ value corrected for the exclusive event selection
\cite{labCDFGamma3JetsCorr}. }
 {\label{fig:effsigma}}
\end{center}

The measured values of $\sigma_{eff}$ from TeV experiments at
Tevetron (CDF and D0) and LHC (ATLAS and CMS) are consistent with
each other within their uncertainties. In the following
calculations, a numerical value $\sigma_{eff}\approx$ 15 mb was used
to estimate the production cross section of $\gamma\gamma$ $+$ 2
jets  from DPS with $\sqrt{s}$ = 13~TeV and 14~TeV at LHC. A number
5 mb was assigned as its uncertainty to estimate its effects on the
final results. The uncertainty on $\sigma_{eff}$ is the dominant
uncertainty for the calculation of the production cross section of
$\gamma\gamma$ $+$ 2 jets from DPS in the following sections

\section{$\sigma_{pp\to\gamma\gamma+2jets}^{DPS}$ calculation}

According to the descripition in above section, the production cross
section of $\gamma\gamma$ $+$ 2 jets from DPS in pp collisions can
be written as

\begin{equation}
\begin{split}
\label{eq2g2jDPS} \sigma_{pp\to\gamma\gamma+2jets}^{DPS} =
\frac{\sigma_{pp\to\gamma\gamma}^{SPS}\times
\sigma_{pp\to2jets}^{SPS}}{\sigma_{eff}} +
\frac{1}{2}\frac{\sigma_{pp\to\gamma+jet}^{SPS}\times
\sigma_{pp\to\gamma+jet}^{SPS}}{\sigma_{eff}}.
\end{split}
\end{equation}

$\gamma\gamma$ production has been calculated at
next-to-leading-order (NLO) some time ago \cite{labDiphotonTJ2000},
supplemented also by gluon initiated subprocesses beyond the leading
order \cite{labDiphotonZL2002} and soft gluon resummation
\cite{labDiphotonRS2000,labDiphotonRS2006}. Recently,
next-to-next-to-leading-order (NNLO) corrections to direct diphoton
production also have become available \cite{labDiphotonNNLO}. The
measurements from LHC
\cite{labCMSDiphoton7TeV,labJTaoDiphoton7TeV,labATLASDiphoton7TeV}
show that the NNLO can give much better agreement with measured data
than the lower order predictions. For the integrated cross section,
the predicted values by NNLO are almost exactly the same
\cite{labATLASDiphoton7TeV} or consistent within the uncertainties
\cite{labCMSDiphoton7TeV,labJTaoDiphoton7TeV} with the measured
ones. So for the production cross sections of $\gamma\gamma$
final-state from SPS, $\sigma_{pp\to\gamma\gamma}^{SPS}$, with
different $\sqrt{s}$ will be obtained from the NNLO calculation with
the package 2$\gamma$NNLO.

For the dijet cross section, the measured data at LHC
\cite{labCMSJets20112012,labATLASJets2011} can be well described by
NLO perturbative QCD (pQCD) calculations from NLOJet++ program
\cite{labNLOJet++} corrected to account for non-perturbative and
electroweak effects. From \cite{labATLASJets2011}, the
non-perturbative correction is within 3\% for jet reconstructed with
the anti-$k_{t}$ clustering algorithm \cite{labAntiKTAlgo} and
distance parameter or cone size R=0.4. The corrections for the
electroweak effect can be negligible if the dijet mass less than
about 1~TeV. In this analysis, the NLO calculations of
$\sigma_{pp\to2jets}^{SPS}$ are performed using NLOJet++ (version
4.1.3) within the framework of the fastNLO package (version 2.3.1)
\cite{labfastNLO}.

NLO pQCD prediction from the program JETPHOX (version 1.3.1)
\cite{labJetPhox} is used for the calculation of $\gamma+jet$ cross
sections from SPS in this paper. This program includes a full NLO
QCD calculation of both the direct-photon and fragmentation
contributions to the cross section. The number of flavours was set
to five. Compared with the measurements of $\gamma+jet$ cross
sections at the LHC, the predictions from JETPHOX multiplied by a
factor close to unity for the corrections of hadronisation and
underlying-event effects give a good description of the
$E_T^{\gamma}$ and $p_T^{jet}$ measured cross sections
\cite{labATLASGJet2010,labCMSGJet2010}.

Different PDF sets are used for the calculations of these three SPS
processes. MSTW2008NNLO \cite{labMSTW2008NNLO} is used for
$\sigma_{pp\to\gamma\gamma}^{SPS}$ calculations with 2$\gamma$NNLO.
CT10NLO \cite{labCT10NLO} is used for both
$\sigma_{pp\to2jets}^{SPS}$ with NLOJet++ in fastNLO package and
$\sigma_{pp\to\gamma+jet}^{SPS}$ with JETPHOX.

The calculations of $\sigma_{pp\to\gamma\gamma}^{SPS}$ are performed
with the factorization and renormalization scales equal to the
invariant mass of two photons, $\mu_{F}$ = $\mu_{R}$ =
$m_{\gamma\gamma}$. The scale uncertainty and PDF uncertainty are
also considered. A simplified and less computationally intensive
estimate of the renormalization ($\mu_{R}$) and factorization
($\mu_{F}$) scale uncertainties is performed by varying these scales
simultaneously by a factor of two up and down around
$m_{\gamma\gamma}$, $\mu_{F}$ = $\mu_{R}$ = 2$m_{\gamma\gamma}$ and
$\mu_{F}$ = $\mu_{R}$ = 0.5$m_{\gamma\gamma}$. 41 eigenvector sets
of MSTW2008NNLO are used to build the PDF uncertainty envelope.

Calculations of $\sigma_{pp\to2jets}^{SPS}$ are derived using
NLOJet++ within the framework of the fastNLO package at a
factorization and renormalization scale equal to the average
transverse momentum ($p_{T}^{ave}$) of the two jets ($\mu_{F}$ =
$\mu_{R}$ = $p_{T}^{ave}$). The uncertainty due to the choice of
factorization and renormalization scales is estimated as the maximum
deviation at the six points ($\mu_{F}$/$\mu$, $\mu_{R}$/$\mu$)=
(0.5, 0.5), (2, 2), (1, 0.5), (1, 2), (0.5, 1), (2, 1) with $\mu$ =
$p_{T}^{ave}$. 52 eigenvector sets of CT10NLO are used to build the
PDF uncertainty envelope.

For NLO calculations of $\sigma_{pp\to\gamma+jet}^{SPS}$ using
JETPHOX, the renormalization, factorization and fragmentation
($\mu_f$) scales are chosen to be photon's transverse momentum,
$\mu_{F}$ = $\mu_{R}$ = $\mu_f$ = $E_{T}^{\gamma}$. Same as above,
52 eigenvector sets of CT10NLO are used to build the PDF uncertainty
envelope.

Above calculations were performed with the strong coupling constant
at two-loop order with $\alpha_s(m_{Z})$ = 0.118 in CT10NLO and
0.117 in MSTW2008NNLO. The uncertainty on $\alpha_s(m_{Z})$ was not
considered in this study. The uncertainty from scales, pdf and
$\alpha_s(m_{Z})$ is around 10\%, 5\% and 1\% respectively
\cite{labATLASDiphoton7TeV,labCMSDiphoton7TeV,labJTaoDiphoton7TeV,labCMSJets20112012,labATLASJets2011,labATLASGJet2010,labCMSGJet2010}.
Compared to the larger uncertainty on the $\sigma_{eff}$ with more
than 30\% used in this study as told in the end of section 2, the
effect on the final results from the uncertainty on
$\alpha_s(m_{Z})$ can be negligible.

\section{Results of $\sigma_{pp\to\gamma\gamma+2jets}^{DPS}$ and expected event rates at LHC}

In this paper, several sets of typical selections at LHC were tried
to calculate the production cross section of $\gamma\gamma$ $+$ 2
jets from DPS in pp collisions,
$\sigma_{pp\to\gamma\gamma+2jets}^{DPS}$. Due to the high level
trigger requirements for $\gamma\gamma$ events at LHC for the higher
energy and higher luminosity collisions, five sets of requirements
on the photons' transverse momentum were considered,
($E_{T}^{\gamma_{1}}$, $E_{T}^{\gamma_{2}}$) $>$ (30, 20)~GeV, (30,
30)~GeV, (40, 20)~GeV, (40, 30)~GeV and (40, 40)~GeV with
$\gamma_{1}$ representing the maximum $E_T$ photon and $\gamma_{2}$
the minimum $E_T$ one of two photons. So for single photon
requirement in the $\gamma+jet$, 3 cases with $E_{T}^{\gamma}>$ (20,
30, 40)~GeV were considered. The photon should be also constrained
in the pseudorapidity region $|\eta|<$2.5. An isolation requirement
is applied on the photon to fulfill the isolation requirement from
experimental measurements
\cite{labATLASDiphoton7TeV,labCMSDiphoton7TeV,labJTaoDiphoton7TeV,labATLASGJet2010,labCMSGJet2010}.
The standard isolation, the $E_T$ sum of partons in a cone of size
$\Delta R$=0.4 around the photon required to less than 5~GeV, is
applied in JETPHOX for the calculation of
$\sigma_{pp\to\gamma+jet}^{SPS}$. For 2$\gamma$NNLO, the smooth
Frixione isolation \cite{labFrixioneISO} on the photons is applied

\begin{equation}
\begin{split}
\label{eqFrixISO} E_{T}^{iso}(\Delta R) < \epsilon \Big(
\frac{1-\cos(\Delta R)}{1-\cos(\Delta R_0)} \Big)^{n}
\end{split}
\end{equation}
where $E_{T}^{iso}(\Delta R)$ is the $E_T$ sum of partons in a cone
of size $\Delta R$, $\Delta R_0$ = 0.4, $\epsilon$ = 5GeV, and $n$ =
0.1. This criterion is found to have the same efficiency as the
standard isolation used for the other generators within a few
percent \cite{labCMSDiphoton7TeV,labJTaoDiphoton7TeV}. Additional
the angular separation between two photons is required to be at
least larger than 0.4 ($\Delta R_{\gamma\gamma}>$0.4) to ensures one
photon will not enter the isolation cone of the other photon, which
is similar to the requirement applied in the data analyses at LHC
experiments ATLAS and CMS
\cite{labATLASDiphoton7TeV,labCMSDiphoton7TeV,labJTaoDiphoton7TeV}.

In this study, jet is reconstructed with the anti-$k_{t}$ clustering
algorithm and cone size R=0.5. Jets are in the acceptance region
with $|\eta^{jet}|<$4.5. Two tries on jet $p_T^{jet}$ were
performed, $p_T^{jet}>$ 20 or 25~GeV. For the dijet events, two jets
should be separated by requiring their angular distance $\Delta
R_{jj}$ greater than 1.0 to avoid the overlapping of two jet cones.
For the $\gamma+jet$ production, the angular distance between
$\gamma$ and jet should be greater than 0.5 ($\Delta R_{\gamma j}>$
0.5) to ensure that the partons belong to the jet will not enter to
the isolation cone of $\gamma$.

Fig.~\ref{fig:XSgg} shows the cross sections of
$\sigma_{pp\to\gamma\gamma}^{SPS}$ computed from the 2$\gamma$NNLO
at $\sqrt{s}$ = 13~TeV and 14~TeV with scales and pdf uncertainties
considered, for different sets of $E_T$ requirements on diphotons.
The scale uncertainty is around 10\% and the pdf uncertainty is
about 4\%. The selection sets in x-axis are the 5 sets of
requirements on ($E_{T}^{\gamma_{1}}$, $E_{T}^{\gamma_{2}}$), number
1 for ($E_{T}^{\gamma_{1}}$, $E_{T}^{\gamma_{2}}$)$>$ (30, 20)~GeV,
2 for ($E_{T}^{\gamma_{1}}$, $E_{T}^{\gamma_{2}}$)$>$ (30, 30)~GeV,
3 for ($E_{T}^{\gamma_{1}}$, $E_{T}^{\gamma_{2}}$)$>$ (40, 20)~GeV,
4 for ($E_{T}^{\gamma_{1}}$, $E_{T}^{\gamma_{2}}$)$>$ (40, 30)~GeV
and 5 for ($E_{T}^{\gamma_{1}}$, $E_{T}^{\gamma_{2}}$)$>$ (40,
40)~GeV. The detailed values can also be found in Table
\ref{tabggXS}. For the central values, the cross section with
$\sqrt{s}$ = 14~TeV is about 9\% higher than that with $\sqrt{s}$ =
13~TeV with the same selection requirements, which is within the
scale and pdf uncertainties.

\begin{center}
\includegraphics[width=7cm]{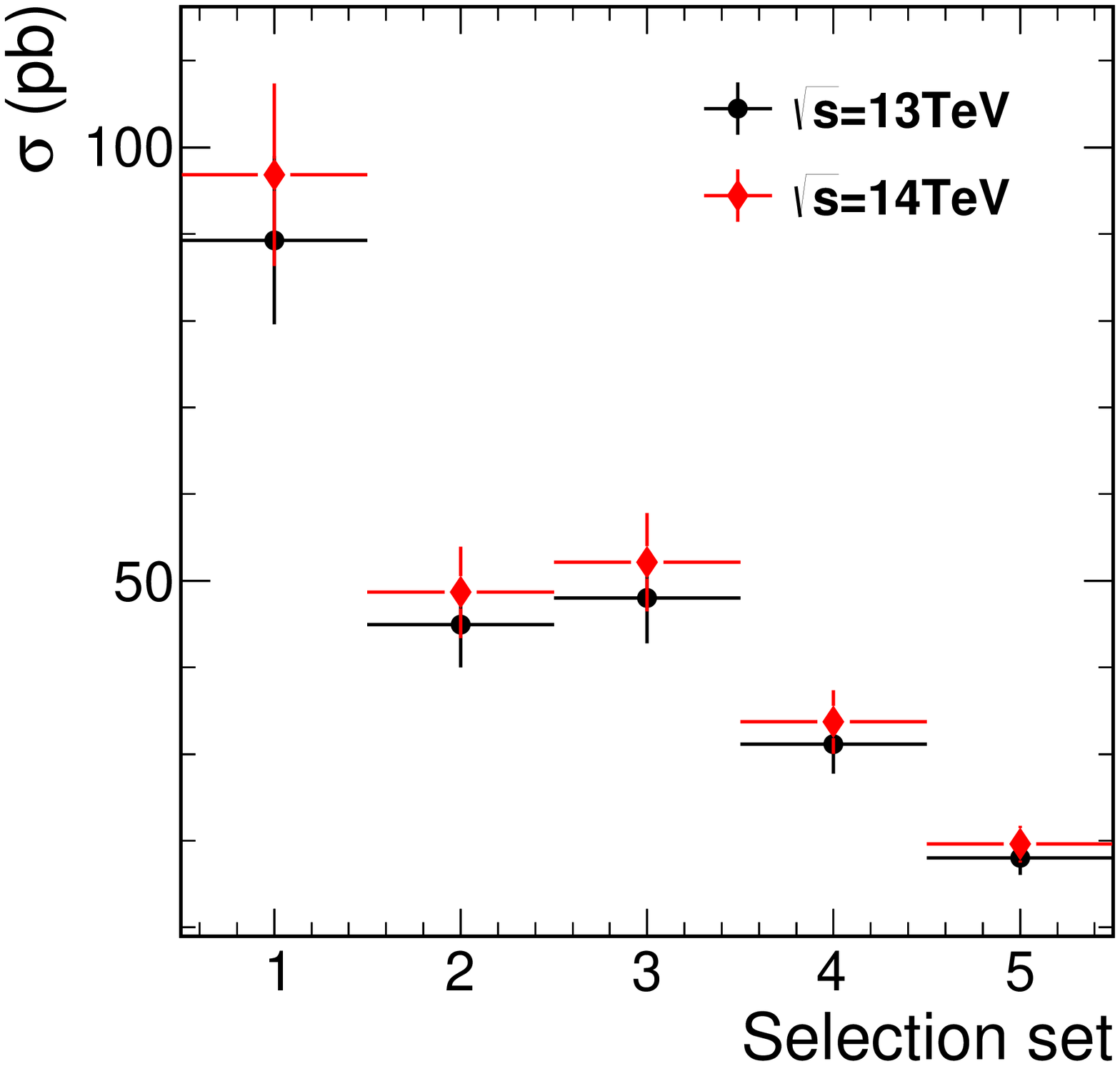}
 \figcaption{ Predicted cross sections of $\gamma\gamma$ by 2$\gamma$ NNLO at $\sqrt{s}$ = 13~TeV and 14~TeV with
 selection set 1 for ($E_{T}^{\gamma_{1}}$, $E_{T}^{\gamma_{2}}$)$>$ (30, 20)~GeV, 2 for
($E_{T}^{\gamma_{1}}$, $E_{T}^{\gamma_{2}}$)$>$ (30, 30)~GeV, 3 for
($E_{T}^{\gamma_{1}}$, $E_{T}^{\gamma_{2}}$)$>$ (40, 20)~GeV, 4 for
($E_{T}^{\gamma_{1}}$, $E_{T}^{\gamma_{2}}$)$>$ (40, 30)~GeV and 5
for ($E_{T}^{\gamma_{1}}$, $E_{T}^{\gamma_{2}}$)$>$ (40, 40)~GeV.
Scale and pdf uncertainties are included.
 }
 {\label{fig:XSgg}}
\end{center}

\begin{center}
\tabcaption{ \label{tabggXS} Cross sections in unit of $pb$ of
$\gamma\gamma$ predicted by 2$\gamma$ NNLO at $\sqrt{s}$ = 13~TeV
and 14~TeV. The uncertainties include the scale and pdf
uncertainties.} \footnotesize
%\begin{tabular*}{80mm}{|c|c|c|}
\begin{tabular}{|c|c|c|}
%\toprule
\hline
 ($E_{T}^{\gamma_{1}}$, $E_{T}^{\gamma_{2}}$)$>$  & $\sqrt{s}$ = 13~TeV  & $\sqrt{s}$ = 14~TeV \\
\hline
(30,20)~GeV & 89.3 $\pm$ 9.7 &  96.9 $\pm$ 10.5 \\
(30,30)~GeV & 44.9 $\pm$ 4.9 &  48.7 $\pm$ 5.3  \\
(40,20)~GeV & 48.0 $\pm$ 5.2 &  52.1 $\pm$ 5.7\\
(40,30)~GeV & 31.2 $\pm$ 3.4 &  33.7 $\pm$ 3.7\\
(40,40)~GeV & 18.0 $\pm$ 2.0 &  19.6 $\pm$ 2.1\\
%\bottomrule
\hline
\end{tabular}
\vspace{0mm}
\end{center}

Fig.~\ref{fig:DiffXSjj} shows the differential cross sections, as a
function of the $p_T$ of leading jet with both jet $p_T^{jet}>$
20~GeV and $|\eta|<$ 4.5, of $\sigma_{pp\to2jets}^{SPS}$ computed
from NLOJet++ within the framework of the fastNLO package at
$\sqrt{s}$ = 13~TeV and 14~TeV with scales and pdf uncertainties
plotted in the same figure. The bottom two plots show the relative
uncertainties including the scale uncertainty and scale$\oplus$pdf
uncertainty combined in quadrature. The contribution of pdf
uncertainty is tiny. The integrated cross section are
117.6$_{-7.0}^{+4.8}$(scale)$_{-1.3}^{+1.0}$(pdf) ($\mu$b) and
122.3$_{-5.1}^{+4.6}$(scale) $_{-1.4}^{+1.1}$(pdf) ($\mu$b) for
$\sqrt{s}$ = 13~TeV and 14~TeV with both jet $p_T^{jet}>$ 20~GeV and
$|\eta^{jet}|<$ 4.5,
52.2$_{-2.4}^{+1.6}$(scale)$_{-0.5}^{+0.4}$(pdf) ($\mu$b) and
56.2$_{-2.3}^{+1.8}$(scale) $_{-0.6}^{+0.5}$(pdf) ($\mu$b) for
$\sqrt{s}$ = 13~TeV and 14~TeV with both jet $p_T^{jet}>$ 25~GeV and
$|\eta^{jet}|<$ 4.5. When $p_{T}$ requirements on both jet increase
5~GeV from 20~GeV to 25~GeV, the cross section are reduced almost by
a factor of 2.

\begin{center}
\includegraphics[width=7cm]{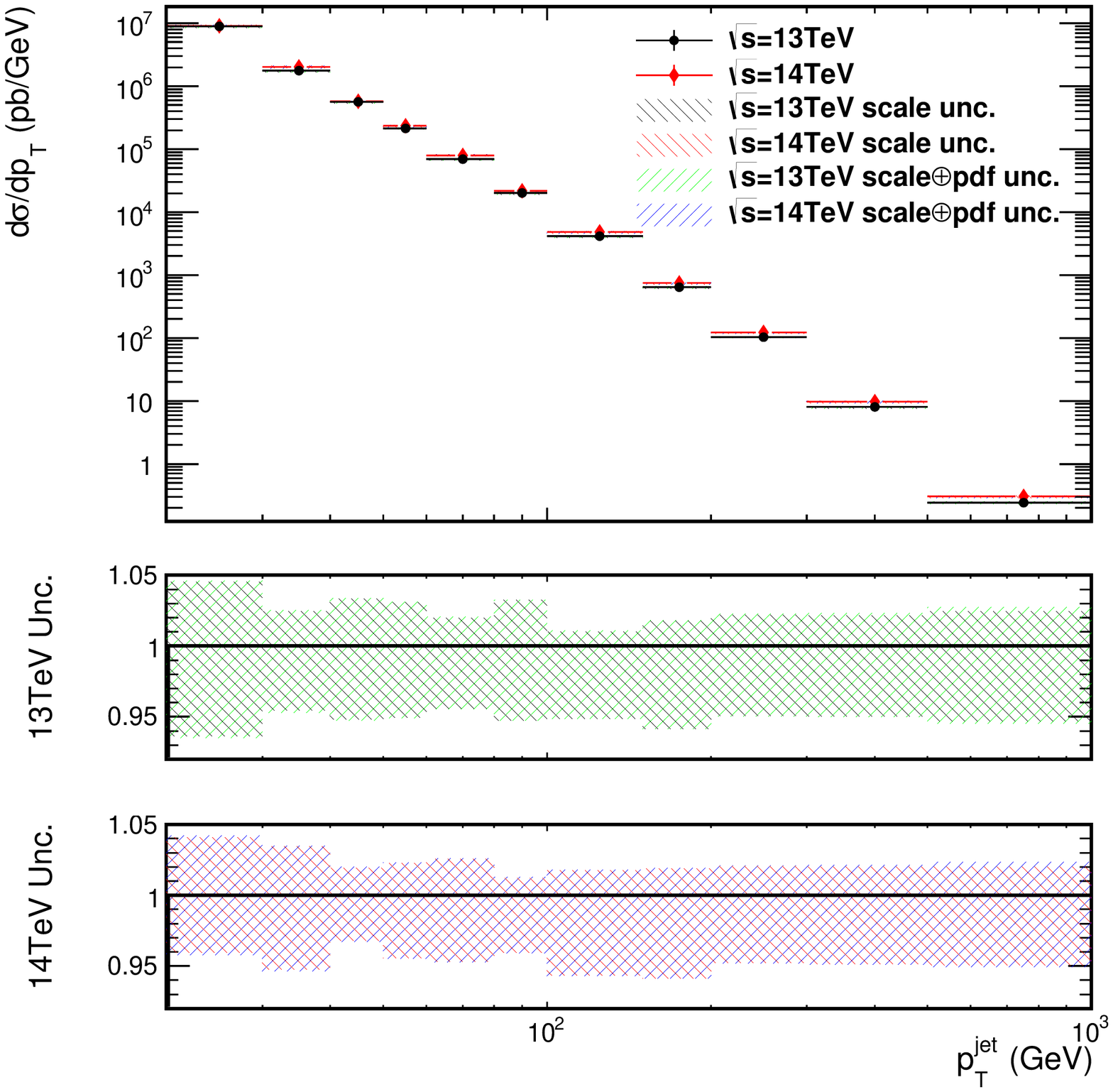}
 \figcaption{ Differential cross sections of $jet+jet$ computed with NLOJet++ within the framework of the fastNLO package,
 after the requirements on both jets with $p_T^{jet}>$
20~GeV and $|\eta^{jet}|<$ 4.5. The solid black circles are the
results for $\sqrt{s}$ = 13~TeV while the red diamond for $\sqrt{s}$
= 14~TeV. Bottom two plots show the relative uncertainties including
the scale uncertainty and scale$\oplus$pdf uncertainty with
$\sqrt{s}$ = 13~TeV in the middle plot and $\sqrt{s}$ = 14~TeV in
the bottom plot.
 }
 {\label{fig:DiffXSjj}}
\end{center}

Combined the photon $E_T^{\gamma}$ requirements for the
$\gamma\gamma$ productions and the jet $p_T^{jet}$ requirements for
the $jet+jet$ productions, cross section of
$\sigma_{pp\to\gamma+jet}^{SPS}$ with six sets of selections on the
transverse momentums of photon and jet with ($E_T^{\gamma}$,
$p_T^{jet}$) $>$ (20, 20)~GeV, (30, 20)~GeV, (40, 20)~GeV, (20,
25)~GeV, (30, 25)~GeV and (40, 25)~GeV were calculated.
Fig.~\ref{fig:DiffXSgj} shows the differential cross sections of
$\sigma_{pp\to\gamma+jet}^{SPS}$ as a function of photon
$E_T^{\gamma}$ with ($E_T^{\gamma}$, $p_T^{jet}$) $>$ (40, 20)~GeV,
$|\eta^{\gamma}|<$ 2.5, $|\eta^{jet}|<$ 4.5 and separation $\Delta
R_{\gamma j}>$ 0.5, computed from JETPHOX with $\sqrt{s}$ = 14~TeV.
The contributions from the direct photon production and photon from
fragmentation are shown in the same plot. The scales and pdf
uncertainties are also plotted in the same figure. The integrated
cross sections are listed in Table \ref{tabgjXS} for different sets
of selections and collision energies. The scale and pdf
uncertainties are also listed in this table, with about 10\%
uncertainty from scales and around 4\% from pdf.

\begin{center}
\includegraphics[width=7cm]{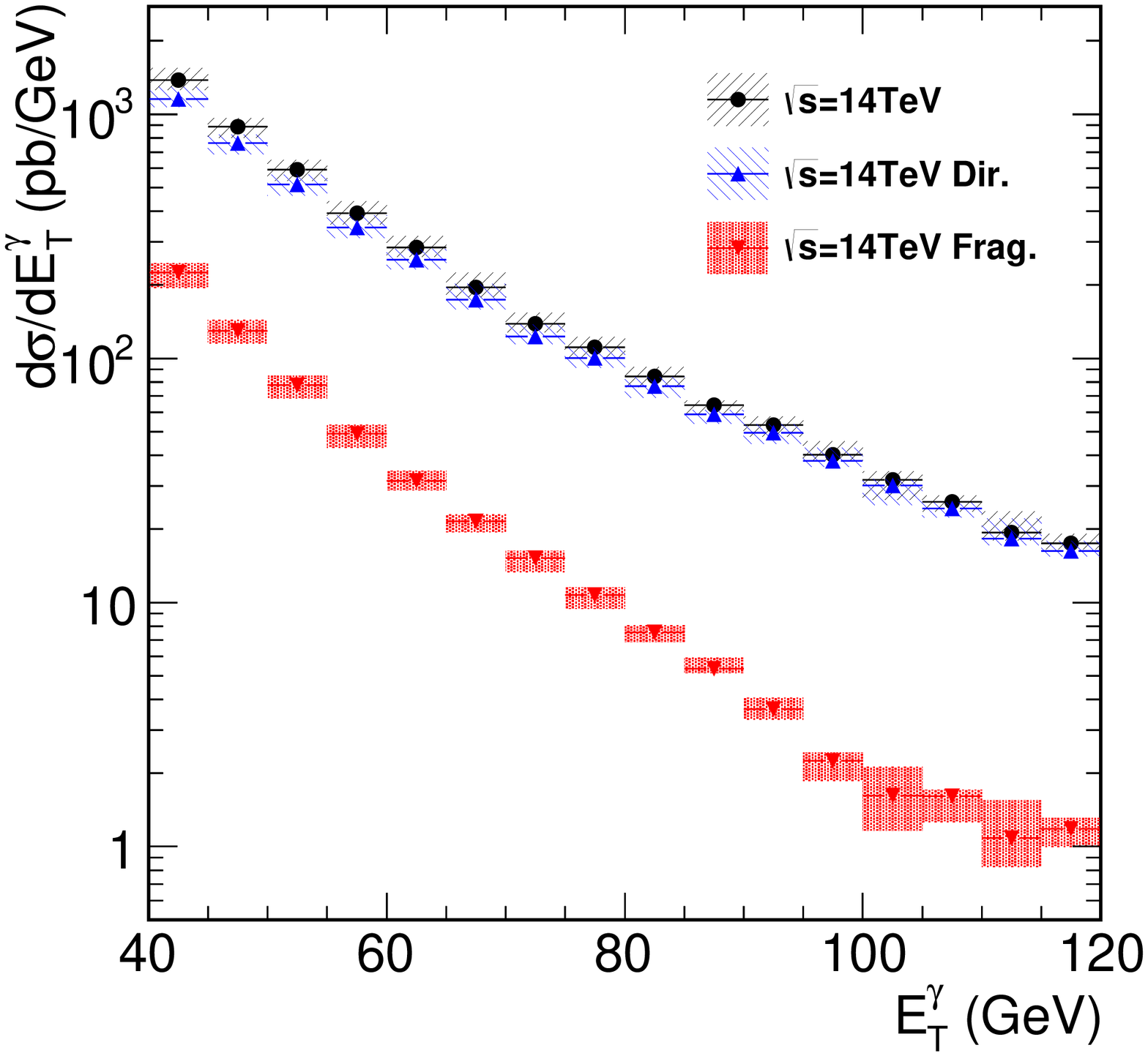}
 \figcaption{ Differential cross sections of $\gamma+jet$ as a function of the photon
$E_T^{\gamma}$, computed with JETPHOX and the selections
($E_T^{\gamma}$, $p_T^{jet}$) $>$ (40, 20)~GeV, $|\eta^{\gamma}|<$
2.5, $|\eta^{jet}|<$ 4.5 and separation $\Delta R_{\gamma j}>$ 0.5
at $\sqrt{s}$ = 14~TeV. The solid circle are the total contributions
while the blue triangles represent the direct contribution and the
red triangles are the contributions of photon from fragmentation.
Scales and pdf uncertainties are also shown in this plot.
 }
 {\label{fig:DiffXSgj}}
\end{center}

\begin{center}
\tabcaption{ \label{tabgjXS} Cross sections in unit of $10^3$ $pb$
of $\gamma+jet$ predicted by JETPHOX at $\sqrt{s}$ = 13~TeV and
14~TeV. The total uncertainties including scale uncertainty and pdf
uncertainty are also list in this table.} \footnotesize
%\begin{tabular*}{80mm}{|c|c|c|}
\begin{tabular}{|c|c|c|}
%\toprule
\hline
($E_T^{\gamma}$, $p_T^{jet}$)$>$  & $\sqrt{s}$ = 13~TeV  & $\sqrt{s}$ = 14~TeV \\
\hline
(20,20)~GeV &  90.1$_{-9.6}^{+11.0}$  &  97.2$_{-11.6}^{+13.3}$ \\
(30,20)~GeV &  48.7$_{-4.6}^{+5.2}$ &  52.7$_{-5.0}^{+5.7}$ \\
(40,20)~GeV &  20.2$_{-1.8}^{+2.2}$ &  22.0$_{-2.1}^{+2.5}$ \\
(20,25)~GeV &  85.0$_{-8.7}^{+9.8}$ &  91.7$_{-9.1}^{+10.9}$ \\
(30,25)~GeV &  41.5$_{-3.9}^{+4.4}$ &  45.0$_{-4.4}^{+4.7}$ \\
(40,25)~GeV &  19.7$_{-1.8}^{+2.0}$ &  21.4$_{-1.9}^{+2.4}$ \\
%\bottomrule
\hline
\end{tabular}
\vspace{0mm}
\end{center}

According to Eq.\ref{eq2g2jDPS} and the above cross sections of the
SPS processes, the cross sections for the production of pairs of
photons plus two additional jets produced from double parton
scattering (DPS) in high-energy proton-proton collisions at the LHC
are calculated for the first time. The results are summarized in
Table \ref{tab2g2jDPSXS}. Two jets in the same
$pp\to\gamma\gamma+2jets$ event from DPS have the same $p_{T}$ cut
thresholds, both $p_{T}^{jet}>$ 20~GeV or 25~GeV simultaneously. The
calculated cross section can be around 0.1 pb to $\approx$1 pb with
the selections considered in this paper. The uncertainty on the
cross section ia around 50\%, with the dominant contribution from
the uncertainty of $\sigma_{eff}$.

\begin{center}
\tabcaption{ \label{tab2g2jDPSXS} Cross sections in unit of $pb$ of
$\sigma_{pp\to\gamma\gamma+2jets}^{DPS}$ calculated for $\sqrt{s}$ =
13~TeV and 14~TeV with the selections described in the paper. The
total uncertainties including scale uncertainty, pdf uncertainty and
also the $\sigma_{eff}$ uncertainty are also list in this table.}
\footnotesize
%\begin{tabular*}{80mm}{|c|c|c|}
\begin{tabular}{|c|c|c|}
%\toprule
\hline
($E_{T}^{\gamma_{1}}$, $E_{T}^{\gamma_{2}}$, both $p_T^{jet}$)$>$  & $\sqrt{s}$ = 13~TeV  & $\sqrt{s}$ = 14~TeV \\
\hline
(30,20,20)~GeV &  0.846$_{-0.432}^{+0.423}$ & 0.960$_{-0.479}^{+0.481}$  \\
(30,20,25)~GeV &  0.428$_{-0.215}^{+0.213}$ & 0.500$_{-0.250}^{+0.250}$  \\
(30,30,20)~GeV &  0.431$_{-0.219}^{+0.215}$ & 0.489$_{-0.242}^{+0.243}$  \\
(30,30,25)~GeV &  0.428$_{-0.215}^{+0.213}$ & 0.250$_{-0.124}^{+0.124}$  \\
(40,20,20)~GeV &  0.437$_{-0.222}^{+0.218}$ & 0.496$_{-0.247}^{+0.247}$  \\
(40,20,25)~GeV &  0.223$_{-0.111}^{+0.113}$ & 0.261$_{-0.129}^{+0.130}$  \\
(40,30,20)~GeV &  0.277$_{-0.141}^{+0.137}$ & 0.313$_{-0.155}^{+0.155}$  \\
(40,30,25)~GeV &  0.136$_{-0.067}^{+0.068}$ & 0.159$_{-0.078}^{+0.078}$  \\
(40,40,20)~GeV &  0.154$_{-0.079}^{+0.076}$ & 0.176$_{-0.087}^{+0.086}$  \\
(40,40,25)~GeV &  0.076$_{-0.037}^{+0.038}$ & 0.089$_{-0.043}^{+0.044}$  \\
%\bottomrule
\hline
\end{tabular}
\vspace{0mm}
\end{center}

With an integrated luminosity of 100 $fb^{-1}$ at $\sqrt{s}$ =
13~TeV accumulated in the following years, about 85k
$pp\to\gamma\gamma+2jets$ events from DPS can be obtained with the
loosest selections, diphoton ($E_{T}^{\gamma_{1}}$,
$E_{T}^{\gamma_{2}}$)$>$ (30, 20)~GeV and both jets $p_{T}^{jet}>$
20~GeV. These events can be triggered by the diphoton paths proposed
at the LHC for $\sqrt{s}$ = 13~TeV. When the integrated luminosity
increasing at $\sqrt{s}$ = 14~TeV, tighter $E_T$ thresholds on
diphoton for the trigger will be used. With the tighter selections,
diphoton ($E_{T}^{\gamma_{1}}$, $E_{T}^{\gamma_{2}}$)$>$ (40,
30)~GeV and both jets $p_{T}^{jet}>$ 20~GeV, about 940k
$pp\to\gamma\gamma+2jets$ events from DPS can be obtained with an
integrated luminosity of 3000 $fb^{-1}$. Even with the tightest
selections studied in this paper, diphoton ($E_{T}^{\gamma_{1}}$,
$E_{T}^{\gamma_{2}}$)$>$ (40, 40)~GeV and both jets $p_{T}^{jet}>$
25~GeV, we can also get about 260k $pp\to\gamma\gamma+2jets$ events
from DPS with 3000 $fb^{-1}$ as designed by LHC.

\section{Summary and outlook}

In this paper, the cross sections for the production of pairs of
photons plus two additional jets produced from double parton
scattering in high-energy proton-proton collisions at the LHC with
$\sqrt{s}$ = 13~TeV and 14~TeV (LHC Run2) are calculated for the
first time. With the generic formula, the cross sections have been
computed based on the theoretical perturbative QCD predictions for
the productions of $\gamma\gamma$ at next-to-next-to-leading-order ,
jet $+$ jet and $\gamma$ $+$ jet at next-to-leading-order, with
their corresponding single-scattering cross sections. From the LHC
measurements with the collision data obtained in years 2011 and 2012
(LHC Run1), these theoretical predictions for these three SPS
processes can give the best agreements with the measured data. With
the typical acceptance and selections used at LHC, the cross
sections $\sigma_{pp\to\gamma\gamma+2jets}^{DPS}$ can be estimated
to be around 0.1 pb to 1 pb with the collision energy $\sqrt{s}$ =
13~TeV or 14~TeV. The expected event rates for $\gamma\gamma$ $+$
2jets from DPS, with some sets of selections, are given for
proton-proton collisions with the collision energy $\sqrt{s}$ =
13~TeV and an integrated luminosity of 100 $fb^{-1}$ planned for the
following two years, and also $\sqrt{s}$ = 14~TeV with 3000
$fb^{-1}$ of integrated luminosity as LHC designed. The
uncertainties on the cross section and the events rates are mainly
dominated by the $\sigma_{eff}$ uncertainty. The scale and pdf
uncertainties for the productions of these three SPS processes are
also considered.

With the incoming LHC Run2 data, there are enough
$pp\to\gamma\gamma+2jets$ events from DPS for investigations. It
needs further studies on the variables, such as the angles between
two photons and two jets, to be chosen for the discrimination of
$pp\to\gamma\gamma+2jets$ events from DPS and
$pp\to\gamma\gamma+2jets$ events from SPS when performing the data
analysis. Also the contributions from the DPS to the whole
$pp\to\gamma\gamma+2jets$ event rates on the distributions of some
typical variables are need detailed investigations in the LHC Run2
data analysis.

\vspace{3mm} \emph{The authors would like to thank Dr. Hua-Sheng
Shao from CERN for helpful discussions.}

\end{multicols}

%\end{linenumbers}

\vspace{-1mm} \centerline{\rule{90mm}{0.1pt}} \vspace{2mm}

\begin{multicols}{2}

\end{multicols}

\vspace{-1mm} \centerline{\rule{90mm}{0.1pt}} \vspace{2mm}

\clearpage

\end{CJK*}
\end{document}